\documentclass[a4paper,11pt]{article}

\usepackage[T1]{fontenc}
\usepackage[utf8]{inputenc}
\usepackage{amsmath}
\usepackage{amssymb}
\usepackage{graphics}

\title{
\textbf{The nature of the narrow peaks
in the $e^+e^-$ pair production 
in heavy ion collisions}
\footnote{Reports of National Academy of Sciences of Ukraine, No12, pp.~91--96 (1998).}
}
\author{P. I. Fomin \and R. I. Kholodov}
\date{1998}

\begin{document}
\maketitle

\begin{abstract}
Experiments on heavy ion collisions performed at GSI (Darmstadt, Germany) revealed series of narrow quasi-equidistant peaks in the energy distribution of the produced $e^+e^-$ pairs in the range from 1.5 to 2.0 MeV which are ascribed to the pair coupling with a strong magnetic field generated by fast nuclei with $Z\sim90$. The pair is captured in the magnetic trap with dimensions of $10^{-10}$~cm and average field  $\sim 5 \cdot 10^{12}$~Gs and, in turn, confines this magnetic field in the trap during $\sim 10^{-19}$~s due to the Alfven effect of magnetic field "freezing-in" into the conducting medium.
\end{abstract}

\section{Introduction}
About a decade ago, in experiments on fast heavy ion collisions performed at GSI (Darmstadt, Germany) the EPOS \cite{one} and the ORANGE \cite{two} groups found a 
series of peaks in the channel of the $e^+e^-$ pair production at the pair energies
\begin{equation}
\label{e1}
1498 \pm 20,\quad 1646\pm 10, \quad 1782\pm 20, \quad 1837\pm 10\mbox{keV}
\end{equation}
in the centre-of-mass frame. Later measurements \cite{three, four} with improved spectrometers confirmed the three latter peaks of the above series (\ref{e1}) and added another peak at $\sim 1575$~keV. Some experiments revealed other peaks located between those mentional above, yet not so easily reproducible.

Both groups \cite{four} believe that the peak reproducibility has to be studied further using improved equipment. However, the nature of these peaks still needs adequate theoretical treatment since numerous attempts to explain the mechanisms underlying the 
$e^+e^-$ resonance formation in the heavy nuclei collisions \cite{five} - \cite{nine} 
have thus far yielded no conclusive results.

This paper presents an essentially new approach to this problem based on the idea to explain the series of quasi-equidistant peaks (\ref{e1}) by the trapping of the $e^+e^-$ pair produced in the nuclear collision, into the magnetic trap formed by a strong magnetic field generated by fast ($v\sim c/9$) heavy nuclei with $z\sim 90$. The key assumption in our approach is that the pair trapped is a relativistic magnetized microplasmoid, which due to the effect of "field freezing - in" \cite{ten}, can confine a rather strong magnetic field in the trap during the finite time $\tau_d$ significantly exceeding the time of flight of the nuclei. The life time of the combined state pair + magnetic field (pairmag) depends on the field diffusion time out of the confinement region which is defined, according to Alfven \cite{ten}, by 
\begin{equation}
\label{e2}
\tau_d=4\pi\frac{\sigma l_c^2}{c^2}
\end{equation}
where $\sigma$ is the plasma conductivity and $l_c$ is the characteristic size of the region.

In the pairmag the electron and the positron wavefunctions are
delocalized within the magnetic trap, so it is natural to expect that the energy
level spectrum of the pair might be similar to that of the Landau levels 
\cite{eleven, twelve} which are determined by a certain 
magnetic field ( $\bar{H}$ ) value averaged over the
confinement region. It will be seen below, that the peaks observed are very well
approximated by the Landau relativistic spectrum \cite{twelve}
\begin{equation}
\label{e3}
E_n=\varepsilon_+  +  \varepsilon_-=
2 \sqrt{
(mc^2)^2+(\pi \frac{c\hbar}{L_z})^2 + 2 c\hbar e \bar{H}\cdot n
}
\end{equation}
with $n_+=n_-=n=0,\, 1,\, 2,\, 3,\, 4,\, 5$ 
for $H=5\cdot 10^{12}$~Gs and $L_c=1.07\cdot 10^{-10}$~cm
where $L_z$ is the longitudinal dimension of the magnetic trap. In this case 
eq.~(\ref{e3}) faithfully reproduces the four lines (\ref{e1}), the $\sim 1575$~keV line and predicts the $\sim 1712$ keV line.

In principle, asymmetric pair states with $n_+ \neq n_-$ must also exist, their
energies lying between the lines where $n_+ = n_-$, but being asymmetric they
might be less probable, i.e. less stable. Nonetheless, these states may play the
role of "intermediate" states in the course of the pairmag evolution leading to its
decay.

\section{Estimation of Basic Pairmag Parameters}

First let us estimate the parameters $L_z$ and 
$\bar{H}$ entering in (\ref{e3}), by making
use of the empirical energies of the first two peaks, on the assumption that
$E_0=1500$~keV and $E_1=1575$~keV. Then we will show that with these $L_z$ and 
$\bar{H}$ eq.~(\ref{e3}) adequately describes the positions of the remaining 
resonances~(\ref{e1}) for $n= 2,\, 4,\, 5$, and for $n = 3$ a new $\sim 1712$
line is found. It is easy to see that
according to (\ref{e3}) the value of $E_o=1500$~keV for $n = 0$ yields
\begin{equation}
\label{e4}
L_z=1.07 \cdot 10^{-10} \quad \mbox{sm}
\end{equation}
and the peak at $E_1=1575$~keV with $n = 1$ and with (\ref{e4}) included gives
\begin{equation}
\label{e5}
\bar H=5 \cdot 10^{12} \quad \mbox{Gs.}
\end{equation}

Substituting the values (\ref{e4}) and (\ref{e5}) into (\ref{e3}) we further obtain
\begin{equation}
\label{e6}
\begin{array}{l}
E_2=1645~\mbox{keV},\quad
E_3=1712~\mbox{keV},\\
E_4=1776~\mbox{keV},\quad
E_5=1839~\mbox{keV}.
\end{array}
\end{equation}

The values of $E_2$, $E_4$ and $E_5$ agree very well with the data (\ref{e1}),
while the $E_3$ line has to be confirmed. 
It might be well to point out that in accordance with this
scheme another line 
\begin{equation}
\label{e7}
E_6 = 1899 \quad \mbox{keV}
\end{equation}
is possible though less probable since as the excitation energy increases, the
pairmag is to be less stable.

By theory \cite{eleven, twelve} the average radius of "Landau's orbits" is defined as
\begin{equation}
\label{e8}
R_H=\sqrt{\frac{c\hbar}{e\bar H}}
\end{equation}
whence for $H = 5 \cdot 10^{12}$~Gs we find
\begin{equation}
\label{e9}
R_H=1.15 \cdot 10^{-10} \quad \mbox{cm.}
\end{equation}

As is evident, $R_H$ and $L_z$ are of the same order of magnitude ($10^{-10}$~sm). These
quantities determine the pairmag size and shape.

It should be noted that colliding heavy nuclei with $Z_{1, 2} \sim 90$ and relative
velocity $v\sim c/9$ are well able to generate a magnetic field of 
$5\cdot 10^{12}$~Gs in the
region of $\sim 10^{-10}$~sm between the nuclei. 
In fact, in this region fields generated by
the two nuclei, add together to produce a total field whose order of magnitude
can be estimated (neglecting the angular dependences) by the formula
\begin{equation}
\label{e10}
\bar H(r)\approx (Z_1+Z_2)\frac{e}{r^2} \cdot \frac{v}{c}, 
\end{equation}
which for $Z_1+Z_2 \sim 180$, \hspace{0.5ex} $v/c \sim 1/9$ can be represented as
\begin{equation}
\label{e11}
\bar H(r)\approx 5\cdot 10^{12}(\frac{r_0}{r})^2 \;\: \mbox{Gs,} \quad
r_0=0.438\cdot 10^{-10} \;\: \mbox{sm.} 
\end{equation}
It is clear that the field averaged over the region of $10^{-10}$~sm can be of the
required order.

In this paper we omit the discussion of the averaging procedure,
introducing the field $\bar H$ as a phenomenological parameter. Note that the
averaging is "performed" by the $e^+e^-$~pair itself whose wave function is
delocalized in the magnetic trap. Moreover, the field in the pairmag may also be
somewhat amplified due to the "stretching" of the lines of magnetic force by the
nuclei flying apart. If one or both nuclei undergo fission, this might influnce the
pairmag parameters and how $e^+$ and $e^-$ would fly apart after the pairmag decay.

\section{Estimation of Pairmag Life Time}
Now we will show that in the pairmag the response ("confining"according to
Alfven \cite{ten}) action of the pair on the magnetic field is also strong, resulting in
the fairly long pairmag life time exceeding by about an order of magnitude the
time of flight of the nuclei across the pairmag area
\begin{equation}
\label{e12}
t_{\mbox{\tiny \textit{fl}}}=\frac{l_c}{v}\sim 9 \: \frac{l_c}{c}
\sim 3\cdot 10^{-20}~\mbox{s,}
\end{equation}
thus providing relatively small peak widths. Note that the time of flight of the
nuclei across the area about the nuclear size is $\sim 3\cdot 10^{-22}$~s.

A criterion for the essential influence of the conducting medium on the
magnetic field dynamics is the large "magnetic Reynolds number" \cite{ten}:
\begin{equation}
\label{e13}
R_m=4\pi \sigma \frac{l_c v_c}{c^2}\gg 1,
\end{equation}
where $v_c$ is the characteristic velocity, $l_c$ is the characteristic dimension, 
and $\sigma$ is the conductivity. 
The criterion (\ref{e13}) is transformed into the Lundquist criterion
for the magnetic field "freezing-in" into the conducting medium \cite{ten}
\begin{equation}
\label{e14}
\Lambda_m=R_m(v_c \! = \! v_A)=
\sqrt{4\pi} \:\; \frac{\sigma l_c \bar H }{c^2\sqrt{\rho}} \gg 1
\end{equation}
if the velocity of magnetohydrodynamic Alfven waves is taken as $v_c$,
\begin{equation}
\label{e15}
v_A=\frac{\bar H}{\sqrt{4\pi\rho}},
\end{equation}
where $\rho$ is the mass density of the medium. 
For $H = 5\cdot 10^{12}$~Gs and $\rho$ determined by 
(\ref{e1}) and (\ref{e3}), the expression (\ref{e2}) for the characteristic 
time of the magnetic field diffusion takes the form
\begin{equation}
\label{e16}
\tau_d=\frac{l_c\Lambda_m}{v_A}.
\end{equation}

Comparing with $t_{\mbox{\tiny \textit{fl}}} = l_c/v$, it is possible to write
\begin{equation}
\label{e17}
\tau_d=\frac{1}{9} \: t_{\mbox{\tiny \textit{fl}}} \, \Lambda_m,
\end{equation}
from which it is seen that the condition $\tau_d\gg t_{\mbox{\tiny \textit{fl}}}$
would hold for $\Lambda_m/9\gg1$.

Next we estimate $\Lambda_m$. Taking into account that $v_A\sim c$, we obtain
$\Lambda_m=4\pi\sigma l_c/c$. We have $l_c\sim 10^{-10}$~sm. 
It now remains to determine the conductivity $\sigma$.
To estimate $\sigma$ in the electron-positron pairmag "plasma" we make use of the
well-known formula \cite{ten}:
\begin{equation}
\label{e18}
\sigma = \frac{e^2 n\tau_{\pm}}{m_e}=\frac{e^2}{m_e v_\pm \sigma_\pm}
\end{equation}
where $\tau_\pm$ is the "time between the $e^+$-$e^-$ collisions", 
$\sigma_\pm$ is the scattering cross-section, and $v_\pm$ is the relative velocity.
The differential cross section for the $d\sigma_\pm(q)$ scattering is 
defined by Bhabha's formula \cite{twelve}. In estimating the integral
cross section for the mutual electron-positron scattering leading eventually to
the pairmag decay, it should be borne in mind that due to the energy levels
being quantized the low-energy (lower than the transition energy between the
adjacent levels) exchanges between $e^+$ and $e^-$ are forbidden:
\begin{equation}
\label{e19}
\varepsilon_n^+ + \varepsilon_n^-=\varepsilon_{n+1}^+ + \varepsilon_{n-1}^-.
\end{equation}
It is seen from the data (\ref{e1}), that the adjacent 
levels for the same particle differ
by approximately $\Delta\varepsilon \sim 35$~keV. 
To perform such transitions $e^+$ and $e^-$ have to
exchange the momentum $\Delta p$ related to $\Delta\varepsilon$
by the expression $\Delta\varepsilon \sim (\Delta p)^2/2m$.
Expressing $\Delta p$ via the initial momentum and the scattering andgle 
$\theta$ as $\Delta p\sim p\theta \sim mc \theta$, for the minimum 
scattering angle we obtain
\begin{equation}
\label{e20}
\theta_{min}\sim \frac{2\Delta\varepsilon}{mc^2}\sim 0.137,
\end{equation}
which roughly corresponds to $\theta_{min}\sim 21^\circ $.

As is evident from (\ref{e1}) and (\ref{e3}), the asymmetric states with 
$n^+ \neq n^-$ are not present in the spectrum observed, 
which might indicate that they are less stable
as compared with the symmetric states. Thus, the transitions 
$(n, n) \rightarrow (n+1, n-1)$ and higher transitions would result
in the short-lived pairmag configurations.
The asymmetric states can be regarded as intermediate states in the transititions
to the decay. Scattering at the angles $\theta > \theta_{min}$ can also
immediately lead to the decay, i.e. to the continuous spectrum. 
Therefore to the total scattering cross section the integral
\begin{equation}
\label{e21}
\sigma_\pm (\theta_{min})=
\int\limits_{\theta_{min}}^{\pi} d\sigma_\pm (\theta)
\end{equation}
would correspond. Bhabha's cross section $d\sigma_\pm(\theta)$ \cite{twelve}
includes the term $1/\sin^4(\theta/2)$ which is likely 
to make the most contribution to the integral (\ref{e21}) in the
vicinity of the lower limit. Retaining only this singular term and expanding
$1/\sin^4(\theta/2)\sim 16/\theta^4$, $\cos(\theta/2)\sim 1$, etc. we would have the 
following cross section asymptotics
\begin{equation}
\label{e22}
d\sigma_\pm (\theta) = 
r_e^2 \: \frac{(1+v^2)(1-v^2)}{v^4}
\frac{d\Omega}{\theta^4}
\end{equation}
where $r_e = e^2/mc^2 = 2.8\cdot 10^{-13}$~sm, $v$ is the $e^+$ and $e^-$ 
velocity in the inertial frame of reference (in terms of $c$),
and $d\Omega = \sin\theta\cdot d\theta d\phi \rightarrow \pi d(\theta^2)$.
The integration of (\ref{e22}) in the limits of (\ref{e21}) yields
\begin{equation}
\label{e23}
\sigma_\pm (\theta_{min}) = 
r_e^2 \: \frac{(1+v^2)(1-v^2)}{v^4}  \frac{1}{\theta_{min}^2}
\approx
7.3 \pi \cdot r_e^2 (1+v^2)(v^{-4}-1)
\end{equation}

For the energies (\ref{e1}) and (\ref{e3}) the average square of the particle 
velocity on Landau's orbits is $v^2 \sim 2/3$ yielding
\begin{equation}
\label{e24}
\sigma_\pm (\theta_{min}) \approx 15.2 \pi \cdot r_e^2
\end{equation}
By substituting this cross section into (\ref{e18}) and (\ref{e14}) and taking
into account that $v_\pm \sim c$, we obtain
\begin{equation}
\label{e25}
\Lambda_m= \frac{4\pi\cdot r_e l_c}{15.2 \pi\cdot r_e^2}\approx 94.
\end{equation}

Substituting $\Lambda_m$ into (\ref{e16}) we find for the pairmag lifetime
\begin{equation}
\label{e26}
\tau_d=10.44 \cdot t_{\mbox{\tiny \textit{fl}}} = 3.13\cdot 10^{-19} \; \mbox{s.}
\end{equation}
The corresponding level width is
\begin{equation}
\label{e27}
\Gamma = \frac{\hbar}{\tau_d}\approx 2 \; \mbox{keV.}
\end{equation}

It is to be noted that this width is considered to be the "natural" level width. The
experimental width might exceed significantly the value (\ref{e27}) owing to the
contributions coming from the instrumental errors, Doppler broadening and
variations in the magnitude of the field confined in the pairmag.

\begin{figure}
\resizebox{\textwidth}{!}{%
\includegraphics{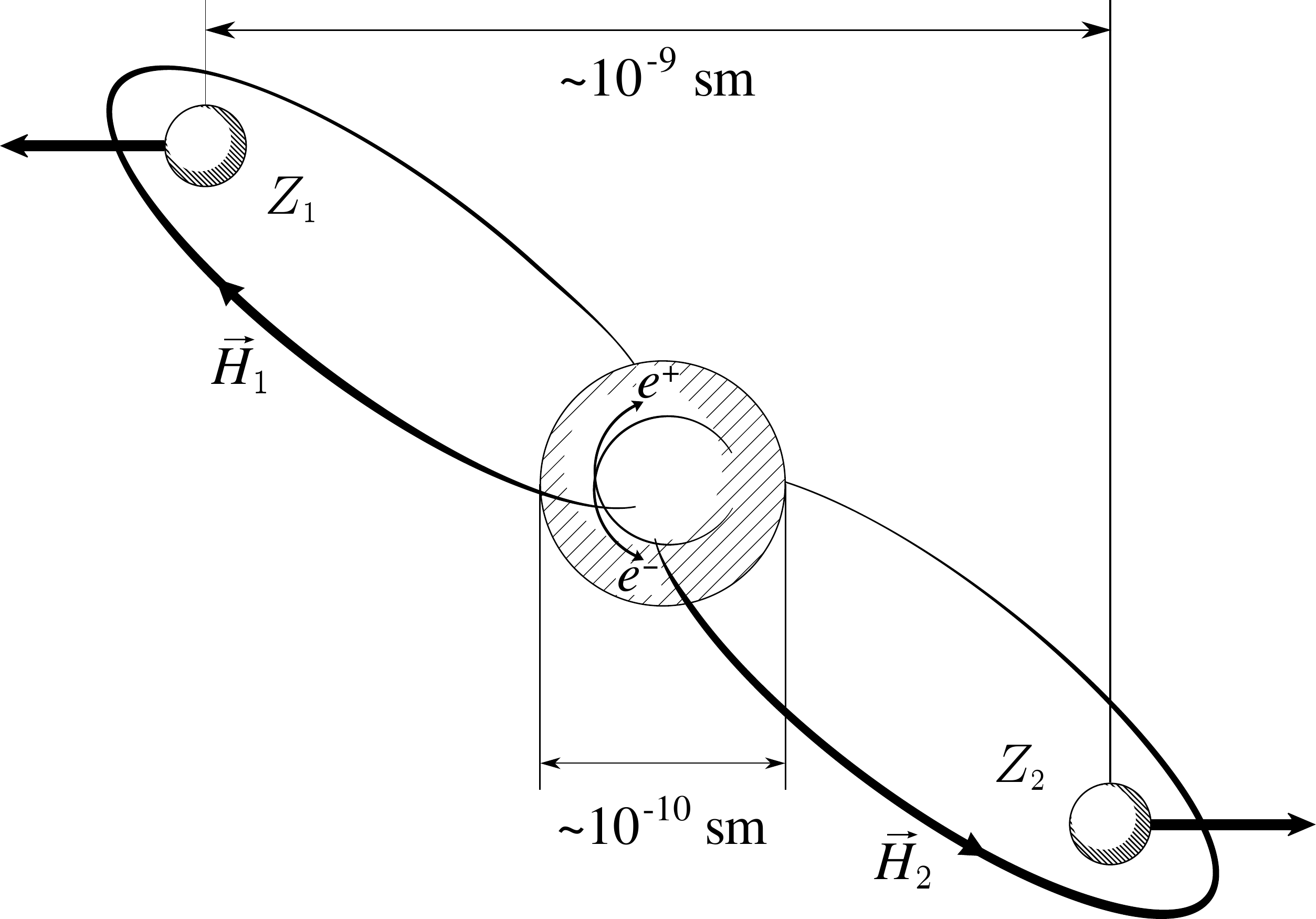}}
\caption{Diagram of the relative positions (in the c.m. frame) of the nuclei
and the pairmag before the pairmag decay.}
\label{fig1}
\end{figure}

During the pairmag life time the nuclei would fly a distance
\begin{equation}
\label{e28}
L_N = \frac{1}{9} \tau_d \: c\approx 10^{-9} \; \mbox{sm}
\end{equation}
apart and half as far away from the pairmag (see fig.\ref{fig1}). This distance is much
greater than the pairmag dimensions. The energy of the Coulomb interaction
between the nuclei and the pairmag at this distance would be an order of
magnitude lower than in the case of the interaction between the pair and the
pairmag magnetic field confined by the pair.

It is worthwhile noting that this field is generated by fast heavy nuclei,
while the $e^+e^-$ pair held in the magnetic trap between the nuclei only confines
the initial field in the vicinity of the pairmag during the time of 
$\sim 10^{-19}$~s due to the Alfven effect of partial magnetic 
field "freezing-in" into the conducting medium.

\section{Conclusions}
As discussed above, the model proposed describes fairly well the
principal characteristics of narrow peaks in the $e^+e^-$ pair production due to the
"pairmag" structure resulting from the trapping of the pair into the magnetic trap
between the nuclei and the confinement of this field in the trap during the time
of $\sim 10^{-19}$~s owing to the high-conductivity "plasma microcloud" formed by the
pair. A number of simplifying semiphenomenological assumptions and
semiclassical estimates were made which are to be replaced later by more
rigorous quantum-electrodynamic calculations. Thus, for example, arguments
have to be given to justify the choice of the pairmag average magnetic field
($\bar H$) of $\sim 5\cdot 10^{12}$~Gs, proceeding from the requirement that the pair
confinement by the field and the field confinement by the pair should be self-
consistent processes. Naturally, the $\bar H$ value can vary slightly from event to
event, contributing to the line broadening. $\bar H$ can be sensitive to the kinematics
of the nuclei flying apart, their possible fission and the way the fission products
would fly apart. All these circumstances are likely to be also important for the
reproducibility of the $e^+e^-$ peaks, making it difficult to provide reliable
experimental data.

In conclusion, the pairmag decay is to be accompanied by emission of a
great number of "quasi-classical" \cite{twelve} soft (up to soft X-rays) photons, whose
total energy may be of the order of 1000~keV. The fact is that not only the $e^+e^-$
pair with Landau-level energies (\ref{e3}), but also the magnetic field itself makes a
significant contribution to the total pairmag energy
\begin{equation}
\label{e29}
E_{\bar H} \sim \frac{\bar H^2}{8\pi} \pi R_H^2 L_z
\sim \frac{c \hbar \bar H}{8e}L_z
\sim 2.6 \;\; \mbox{MeV.}
\end{equation}

Since the initial field $\bar H$ is created by moving nuclei, 
the $\bar H$ energy is partially
transferred to the nuclei flying apart and partially radiated. The calculations of
the spectrum and angular distribution of the radiation go beyond the scope of
the present paper. We would only note that the response action of the soft
radiation on the $e^+e^-$ pair fragments flying apart 
might be significant in the case of the asymmetric distribution.

The authors wish to thank V.~E.~Storizhko, V.~I.~Miroshnichenko, \\
S.~P.~Roshchupkin for many useful discussions and V.~P.~Gusynin for valuable
original information. The work was supported in part by the State Committee
for Science and Technology Fund, Grant No109 and the STCU Foundation,
Grant No2.4./586.


\end{document}